\documentclass[11pt,twoside]{article}
\usepackage{asp2004}\usepackage{psfig}\usepackage{epsf}\usepackage{graphics}\usepackage{lscape}
\makeindex
\begin{document}
\title{Pro-Am Collaboration and the AAVSO}
\author{A. A. Henden}
\affil{AAVSO, 25 Birch St., Cambridge, MA 02138 USA}

\begin{abstract} Professionals need to be aware that there is a
valuable resource available and waiting to be used - the amateur
astronomy community.  We give some examples of how pro-am
collaborations have worked in the past, indicate the advantages
and disadvantages of such collaborations, and suggest methods
by which a professional can find and work effectively with
amateur astronomers.
\end{abstract}

\section{Introduction\label{sec:intro}}

There is a resource
of which all professionals should be aware: the opportunity for
professional-amateur (pro-am) collaboration.  This can be as simple
as working with a single, local amateur, or as complex as involving
a large group on a project for a number of years.

There is a long history of pro-am interaction.
In fact, the first professionals were hired by amateurs to
maintain their observatories and help in observing and
data reduction.  An example of this is the 
Lowell Observatory (Flagstaff, Arizona),
founded in 1894 by Percival Lowell, a rich Bostonian.
He hired V. M. Slipher to maintain the observatory during
those periods when Lowell was not in Flagstaff; Slipher of
course used the 60cm Alvan Clark refractor plus a Brashear
spectrograph to measure redshifts of galaxies.

It has only been in the past few decades that amateurs have
stood in their own right.  It is primarily due to the
advent of cheap, high-quality instrumentation, plus the
influx of technologically proficient observers.
The traditional separation between the professional and
the amateur is becoming blurred.

During the first half of the 20th century, most amateur
observations were made visually, using small aperture
telescopes.  They concentrated on the long-period,
high-amplitude variables, but lacking a strong connection
to the professional community.
The Space Age called many amateurs into duty to monitor
artificial satellites during the Smithsonian Astrophysical
Observatory's Moonwatch program. This program is
described in Cornell (1975), and lasted for nearly 20 years
with 400,000 observations of nearly 6,000
artificial satellites.

A major change after WWII was the release
of the 1P21 photomultiplier tube.  War surplus
equipment plus a large base of militarily trained,
electronically proficient, amateurs led to the
development of highly precise photoelectic photometers.
Wood (1963) wrote a book
on photoelectric photometry (PEP) for amateurs, and
PEP committees were formed
in several of the larger amateur organizations.
The photoelectric photometer gave amateurs the
capability of making observations as precise as
the professional, albeit on brighter objects.

In 1980, the International Amateur Professional
Photoelectric Photometry (IAPPP) organization was
founded by R. Genet and D. Hall.  This was an
active organization, promoting photoelectric
astronomy and pro-am collaboration.

Several papers have been written about pro-am
collaboration. 
Hearnshaw and Cottrell (1986) held an IAU colloquium
on instrumentation and research programs for small
telescopes.
 Henbest (1987)
describes a joint meeting between the British
Astronomical Association, an amateur group, and
the Royal Astronomical Society, a professional
organization, held in March of 1987.  The summary
indicated a desire on the part of professionals
for closer collaboration on many projects,
including monitoring Saturn for new cloud
features, discovering comets, measuring the
position of asteroids, and further photoelectric
observations of just about everything.
Percy et al. (1992)
edited the proceedings of a meeting on International
Cooperation and Coordination in Variable Star
Research, attended by professionals and
amateurs alike.
Smith (1995) and others wrote papers for the
Journal of the British Astronomical Association
on the Centenary Meeting of the BAA variable
star section, describing their experience in
pro-am collaborations..
Millis (1996) organized a workshop at Lowell
Observatory on the Role of Small Telescopes in
Modern Astronomy, highlighting the critical role
of small telescopes, often run by amateurs, in
many areas of research.
Percy and Wilson (2000) held a conference on
amateur-professional partnerships in astronomy,
highlighting many collaborative areas.
The American Astronomical Society (AAS) recognized
the importance of pro-am projects
through the formation of the Working Group on
Professional-Amateur Collaboration, a standing
committee in the AAS with its own web site.
Finally, Price (2005) discusses new areas of
potential pro-am collaboration such as datamining.

Pro-am collaboration crosses the entire face of astronomy,
from sunspots to gamma-ray afterglows. Talk to anyone
working in astronomy and often an early interest in amateur
astronomy will be found.
To discuss this pro-am resource in the the specific context of
variable-star research, the American Association of Variable
Star Observers (AAVSO) will be used as an example.

\section{The American Association of Variable Star Observers (AAVSO)}

\begin{figure}[!ht]
\centerline{\hbox{\psfig{figure=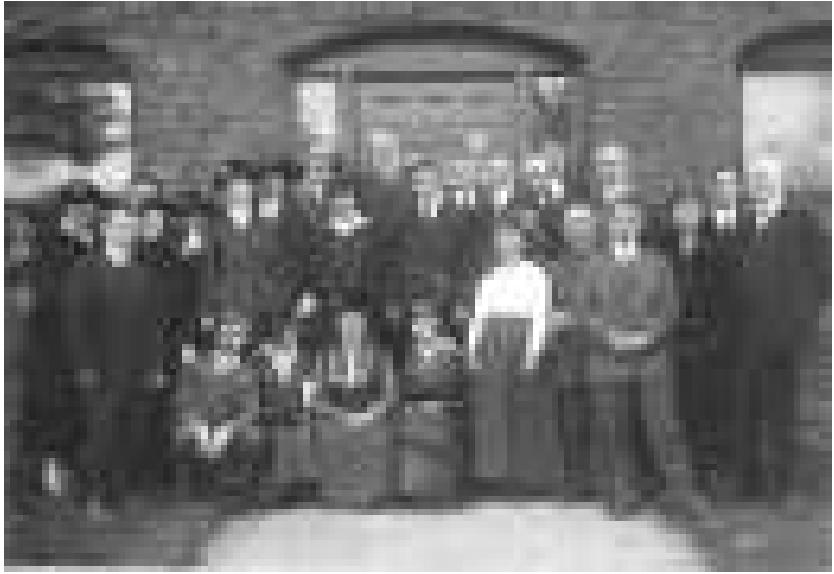,angle=0,clip=,width=11cm}}}
\caption[]{AAVSO Meeting, Harvard 1917}
\label{fig:meet}  
\end{figure}

The AAVSO was established in 1911 by William Tyler Olcott,
an avid amateur in the Boston area.  The initial organization
was closely associated with the Harvard College Observatory (HCO).
In fact, the inital steps towards incorporation
were taken at the request of E. C. Pickering, the
Director of the HCO, who offered
space and technical guidance.  The first formal
meeting of the AAVSO took place in 1917 and is shown in
Fig.~\ref{fig:meet}.

\begin{figure}[!ht]
\centerline{\hbox{\psfig{figure=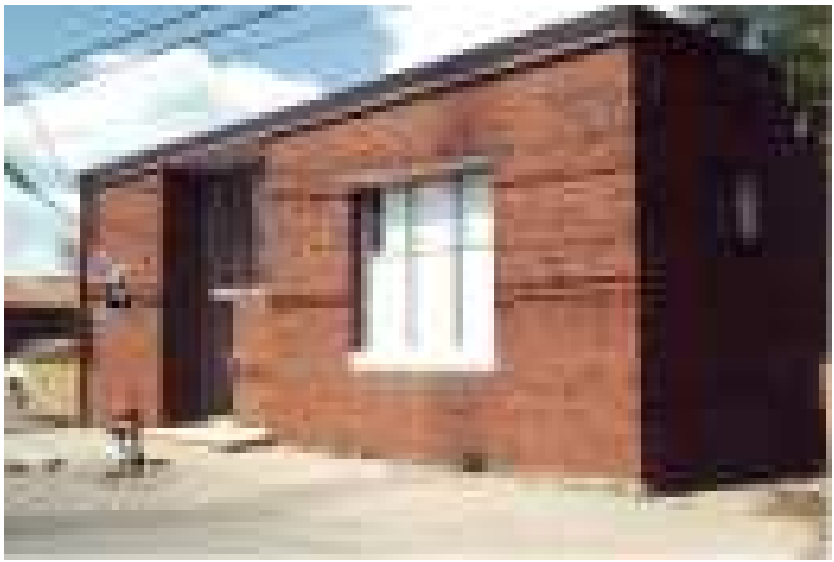,angle=0,clip=,width=11cm}}}
\caption[]{AAVSO Birch Street Headquarters building}
\label{fig:hq}  
\end{figure}

The AAVSO was originally affiliated with HCO,
having an office on the Observatory grounds.
Around 1954, it split off and became a private non-profit scientific
organization.  In 1986, the AAVSO purchased its current
headquarters building in Cambridge, MA (Fig.~\ref{fig:hq}),
about 2km from the HCO.  The AAVSO employs a permanent
staff of 10, with several additional part-time and contract
employees.  The majority of the funding is through a private
endowment, but the AAVSO does aggressively pursue funding via
external research and education grants.

The AAVSO is dedicated to the study of celestial objects that
vary in optical brightness.  They focus efforts in three directions:
education, motivating and mediating professional-amateur astronomical
collaborations, and data archiving. Variable-star astronomy is
used to teach universal science concepts and get children and
the public addicted to science.  The AAVSO works to
bring together amateur observers with  professional astronomers
who need data and need research assistance, allowing greater science
to be done at a lower cost by more people. It also provides
a single location where data from thousands of observers and
observatories can be stored and searched, creating a virtual data
clearing house for scientists the world over.

The AAVSO is one of the largest pro-am astronomy
organizations in the world.  It has 1200 members in 40 countries;
about 20\% are professionals.  Another 2000 non-member
observers and groups in 60 countries submit observations annually.
Approximately 900,000 observations are archived annually, with
a total International Database of 13 million observations, some
dating back over 150 years.  It maintains a permanent website
at http://www.aavso.org where observations, charts, sequences,
tutorials and other materials can be accessed.

The AAVSO is not the only amateur organization, though it is the
largest such organization devoted to the study of variable
stars.  Others include: the Variable Star Observers League
in Japan (VSOLJ; it grew from the Oriental Astronomical Association,
founded 1920), the Association Francaise de
Observateurs d'Etoiles Variables (AFOEV; founded 1921),
and the Bundesdeutsche Arbeitsgemeinshaft
f\"{u}r Ver\"{a}nderliche Sterne (BAV; founded 1950);
and the variable star sections of many large amateur organizations
such as the British Astronomical Association (BAA; founded 1890)
 and the Royal Astronomical
Society of New Zealand (RASNZ; founded 1920).  More information
on any of these organizations can be found on the Web.

There are possible reasons to prefer working
with a variable-star organization other than the AAVSO.
Having amateurs physically nearby means that personal
mentoring is possible, as well as visits to meetings to
present research projects.  Local amateurs will often become
frequent collaborators.  Nearby clubs will also speak a
common native language, useful for communication.  While
English is becoming the de facto scientific language, not
all amateurs can communicate effectively in it.  Local
clubs and affiliates of larger national organizations are
also good sources of new observers, as most amateurs tend
to join an astronomical organization as a stepping stone
to more advanced observing.

\section{Telescopes}
\begin{figure}[!ht]
\centerline{\hbox{\psfig{figure=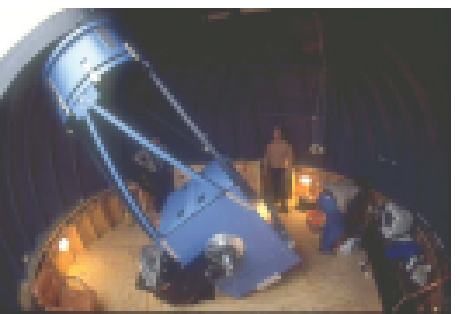,angle=0,clip=,height=6cm}\hspace{3mm}\psfig{figure=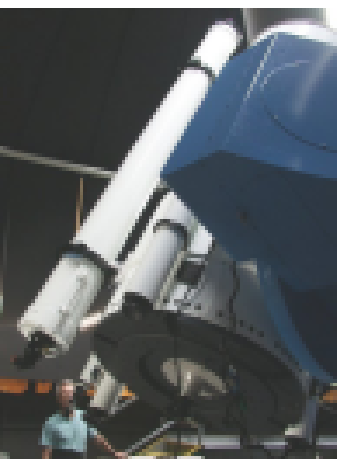,angle=0,clip=,height=6cm}}}
\caption[]{{\sl Left:} Motta Observatory. {\sl Right:} Georges Observatory.} 
\label{fig:motta}  
\end{figure}

One common assumption is that all amateur telescopes are
small and not useful for scientific projects.  While it is
true that most amateurs do not have a 4m-class telescope
in their backyards, many of the observatories are the
equal of university facilities.  Fig.~\ref{fig:motta}a shows the 80cm telescope
of M. Motta in Massachusetts; there are many amateurs with
this class telescope.  Even more common are amateurs
associated with planetariums and public observatories.
Fig.~\ref{fig:motta}b shows the 90cm telescope of Georges Observatory,
Ft. Bend, Texas; a facility that is used for public
viewing on a few nights per year, but available to club members
the rest of the time.  The Faulkes 2m telescopes are also
available to educational institutions; the educators are often
associated with amateurs or are amateurs themselves.

Even if $ 0.3-0.5$m telescopes are considered, with modern
CCDs these telescopes are fully capable of providing
time-series photometry at moderately faint magnitudes.
For spectroscopic support, these mesh well with 4m-class
telescopes. The quality of modern telescopes is quite
amazing, with small periodic error and mount hysteresis.
With built-in autoguiding, long exposures and accurately
registered long time series are possible.  Field
misidentification is a thing of the past.  Most modern
telescopes are GOTO, meaning a flexure map is in
place and giving the telescope computer coordinates
will position the telescope to within an arcmin or so.
Software is available to fully automate a telescope
plus camera system.  There
are far more fully robotic amateur systems today than
exist in the professional world.

At the same time, many professional surveys are using
smaller telescopes, such as the 10cm telescopes used
by the TrES project (Alonso et al. 2004).  For these
systems, the amateur telescope may be the larger,
followup observatory.  This is another example of
how the line between professional and amateur has
been nearly erased.

\section{Visual Observations}

\begin{figure}[!ht]
\centerline{\hbox{\psfig{figure=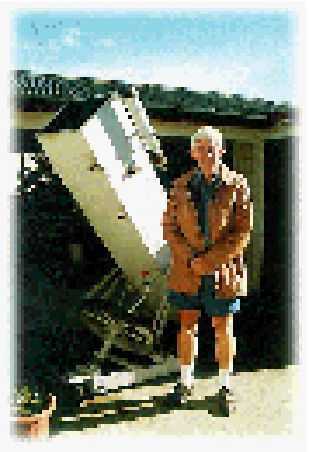,angle=0,clip=,height=6cm}\hspace{3mm}\psfig{figure=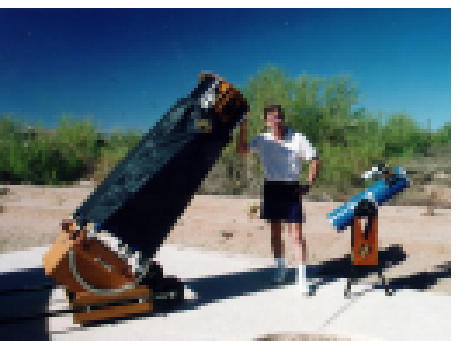,angle=0,clip=,height=6cm}}}
\caption[]{{\sl Left:} Albert Jones.   {\sl Right:}  Gene Hanson.} 
\label{fig:jones}  
\end{figure}

It is certainly true that visual observations formed the
early history of amateur variable-star astronomy.
Fig.~\ref{fig:jones}a is
of A. Jones, a prolific observer from New Zealand,
who designed and built his 32cm reflector nearly 60 years
ago.  He has made an estimated half-million observations over
his career.  Fig.~\ref{fig:jones}b is of G. Hanson, an Arizona amateur,
who with his 45cm Dobsonian telescope has made over
50,000 observations. This
is the traditional appearance of a visual amateur, much
as the public's perception of a professional astronomer is like
the photographs of Percival Lowell at the eyepiece of
his 60cm Alvan Clark refractor.

For visual observers, the major change has been the
availability of inexpensive, high quality telescopes.
Hanson's Dobsonian is a large-aperture, easily positioned
telescope that permits him to make more than a hundred
estimates on any given night.  Vendors such as Meade
and Celestron are making Schmidt-Cassegrain telescopes
with GOTO features for rapid acquisition of
new fields.  An ancillary change has been the creation
of better finding charts, providing improved photometry
for the comparison-star sequence.  Better sequences enable
the observer to reach his/her potential of about 0.1-0.2mag
error per estimate.

\begin{figure}[!ht]
\centerline{\hbox{\psfig{figure=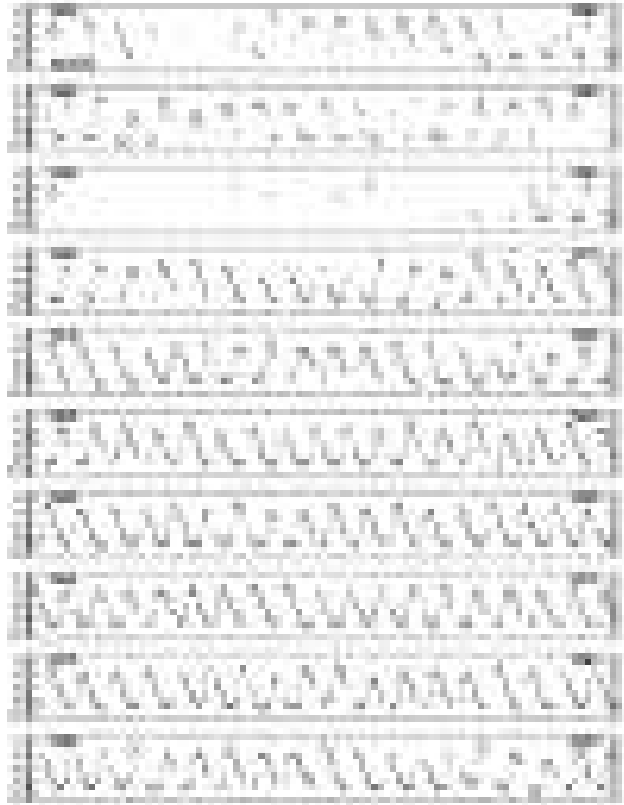,angle=0,clip=,width=9cm}}}
\caption[]{Light curve of Mira.}
\label{fig:mira}  
\end{figure}

The advantage of visual observing is that, because minimal
equipment is involved, the process has been ongoing for
over a century.  A typical light curve for Mira,
the prototype long period variable, is shown in Fig.~\ref{fig:mira}.
This covers the period 1850 through 2000, showing the changing
aspect of the light curve.  Note that there
are gaps in the data before about 1890; the AAVSO is
searching for early datasets that have not been added
to our database.  Many early observers kept their data
in personal logbooks, and often never submitting them to
any formal organization.  An example of this are the
250,000 observations taken during the early 20th century
by A. W. Roberts, an observer in South Africa, that
we have recently been able to find, digitize and add to our database.

The long monitoring data on many of the brighter variables,
especially long-period variables, has been invaluable for
many secular evolutionary studies.  Wood and Zarro (1981)
studied 3 Miras and compared the observed period changes
with theoretical models.  Templeton et al. (2005)
used the hundreds of Mira variables in the AAVSO database
to search for period changes.  Bedding et al. (2002) studied
the semiregular variable $L_2$ Puppis, showing that it
has had a decade-long dimming event from dust.  These are
all studies that would be impossible without the long
observational record from the visual observers.

\begin{figure}[!ht]
\centerline{\hbox{\psfig{figure=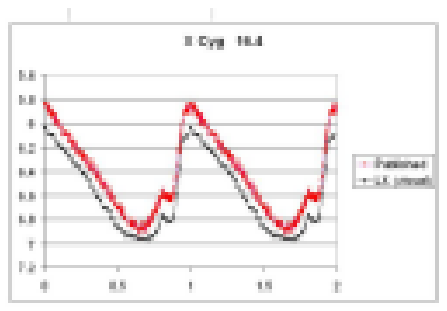,angle=0,clip=,width=11cm}}}
\caption[]{Light curve of X Cyg.}
\label{fig:xcyg}  
\end{figure}

Professionals also often complain about the precision
of visual observers.  True, each individual observation
has an imprecision of 0.1-0.2mag for the better observers,
but as long as the errors are random, averaging multiple
observations can dramatically improve the light curve.
Two examples of this are shown here.  Fig.~\ref{fig:xcyg} shows the
cepheid variable X Cyg, where the data from one visual
observer, but from multiple pulsational cycles, is period-phased
and phase-interval averaging is used.  This visual curve is shown
underneath the curve generated from published photoelectric
data.  Note that small bumps with height less than 0.1mag
are readily apparent in both curves.  This kind of
averaging works well for stars that faithfully repeat
their light variation from cycle to cycle.

\begin{figure}[!ht]
\centerline{\hbox{\psfig{figure=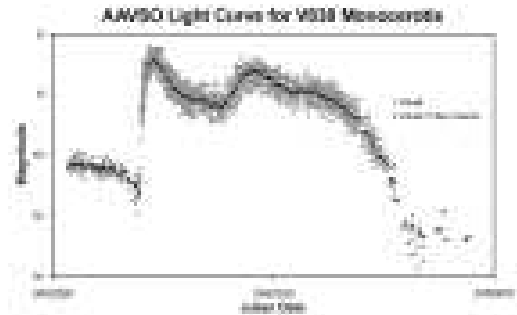,angle=0,clip=,height=4.3cm}}}
\caption[]{Visual light curve of V838 Mon.} 
\label{fig:v838}  
\end{figure}

For transient or non-repeating objects, the peculiar novae V838 Mon is given 
as an example.  Fig.~\ref{fig:v838} shows the visual light curve for the 
variable using one-day means of the individual measures.  There is a small
zeropoint shift between the photoelectric and the visual
data, due to the extreme red color of this nova. Such
light-curve improvement is only possible when the object
is well-observed, and data from many visual observers is combined.

\begin{figure}[!ht]
\centerline{\hbox{\psfig{figure=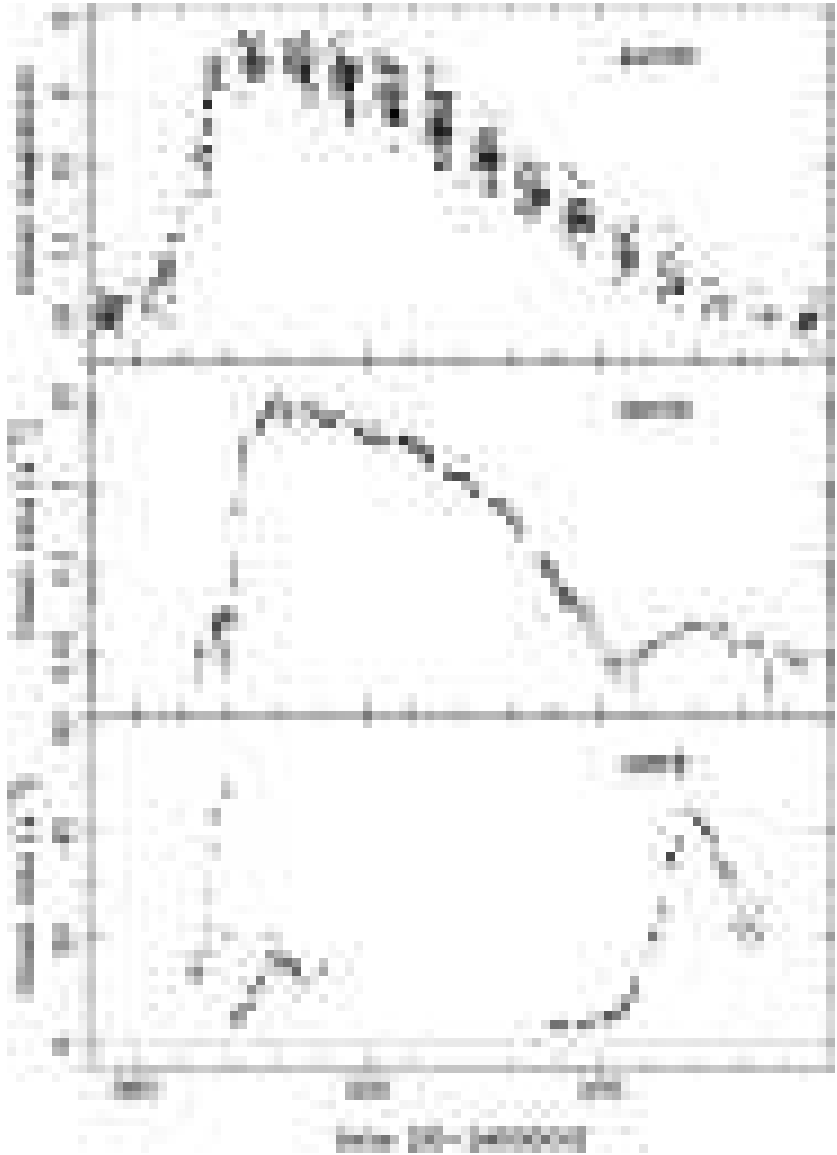,angle=0,clip=,width=11cm}}}
\caption[]{SS Cyg multiwavelength comparison. From Wheatley et al. (2003)}
\label{fig:sscyg}  
\end{figure}

Another modern use for visual data is for determining the
start of an outburst for triggering Target of Opportunity
observations at large ground-based telescopes and space-based
missions.  Fig.~\ref{fig:sscyg} is from Wheatley et al. (2003),
showing a comparison between the AAVSO visual data and
the simultaneous EUVE and RXTE satellite data for an
outburst of the cataclysmic variable SS Cyg.  The visual
observers monitored the CV, and triggered the satellite
observations when the start of an outburst was detected.
The simultaneous observations show that the optical
outburst preceded the EUV outburst by 0.6day and the
X-ray outburst by 0.9-1.4day.  The subsequent rise/fall
pattern supports the view that both hot components arise
in the boundary layer between the accretion disc and
the white dwarf surface.  The opposite case can also be
of importance.  A recent HST program on recently discovered
CVs by Szkody et al. (2005) required that the CVs had to
be in quiescence for instrument safety.  Amateur monitoring
confirmed the quiescent behavior in the hours prior to the
HST observation.

\section{Photographic Progress}

While amateurs have always been interested in astronomical
photography, little photometric work using photographic
plates or film has been done by amateurs.  The photographic
process is inherently non-linear, making photometry difficult.
Usually the results are not much better than visual
techniques, though adequate for discovering and measuring
high-amplitude variables.  See Kaiser (1995) for a typical
program.  However, with modern 35mm cameras, imaging large areas
of the sky looking for transient objects is still viable.
Several novae surveys are underway in Japan and Chile
using photographic techniques. An example is the survey
by Liller (1999).
The larger use of photography is actually archival
searches of professional surveys, as described later.

\section{Modern Photometric Equipment}

Not all amateurs are visual observers.  Over half of
those currently joining the AAVSO have some
sort of CCD camera or electronic instrumentation.
This fraction will only increase in the years ahead.

While some amateurs built their own single-channel
photoelectric photometer after WWII, others purchased
instruments available commercially in the 1970's and
1980's.  In particular, the development of the solid
state photometer such as the Optec SSP-3 brought
many amateurs into the PEP fold.  This was a simple,
small instrument suitable for the back-end weight capacity
of amateur telescopes, and required little instruction
to make work effectively.  It covered the entire
silicon spectral response, with the two drawbacks that
it was a manual instrument (viewing eyepiece, flip
mirror, non-motorized filter slide) and that it was
only effective on relatively bright stars.  Still,
there were hundreds of members in the IAPPP in its
heyday, contributing valuable multifilter photometry
to many scientific projects.

The real change occurred in the 1990's, when CCD
technology advanced to the point that inexpensive
cameras could be constructed.  Berry et al. (1994)
created the Cookbook CCD camera that
was eventually used by over 2500 amateurs.
Commercial cameras were available from Spectrosource, and eventually
others such as Santa Barbara Instrument Group (SBIG) and Starlight
Xpress.  Once there was a sufficient
market, sophisticated camera control and image
processing software became available as well.
The commercial market changes rapidly, so new CCDs
from chip vendors are incorporated quickly, along
with hardware improvements such as USB image download.
In fact, many
amateur cameras are more capable today than their
professional counterparts.

The low-end market is developing even faster.
The Meade DSI-Pro CCD camera is currently
available on the commercial market for about USD\$400.
Other cameras have been announced in the under
USD\$1000 market, and they all show definite potential
in providing accurate magnitude estimates.  These
lower-cost cameras are opening up the field to
smaller telescopes and those in third world countries,
often in an underrepresented geographical location.

Other innovations in the commercial market have been
the tip/tilt low-order adaptive optics modules,
spectrographs of varying resolution, and autoguiders.
The SBIG spectrographs in particular are being
used by many amateurs, obtaining spectra of the
brighter stars and novae that are excellent.
There are several groups building spectrographs
as a team effort.  Spectroscopy is not relegated
to low resolution; some groups (such as Spectrashift)
are working with bench-mounted spectrographs and
obtainng 100m/s doppler measurements of exoplanetary
systems.  A book by Tonkin (2002) on amateur
spectroscopy is available.

\section{Archival Research}

As mentioned earlier, amateurs can visit existing
plate archives and examine old photographic plates,
looking for times of minima, outbursting behavior
or progenitors of unusual objects.  An example of how
such plate archives are used is given in Turner (2003).
There are several
large plate archives in existence.  Perhaps the best
known is the Harvard Plate Collection, currently maintained
by A. Doane.  This archive contains about a half-million
plates taken in the early 20th century.

Plate archives seldom exist in the city where a professional
lives who is working on a research project.  Glass plates are
rarely shipped anywhere due to their fragile nature. However, there
are almost always amateurs near the archive, and in
addition, amateurs that both know how to use the archive
and that have permission to do so.  Others are willing
to travel to the appropriate city.  This is especially
true for the Harvard collection, since the AAVSO Fall
meeting is always held near Harvard and amateur researchers
often add a day or two to their travels to search the
stacks.

Archival plate searches are often the only method by which
progenitor activity can be studied for peculiar objects.
Goranskij et al. (2004) inspected the archival plates
at Sonneberg Obervatory and the Sternberg Astronomical Insitute
for V838 Mon, finding that the B-type companion star seen in recent
spectra dominated the precursor light for this unusual
nova.

The professional community has seen the scientific value
of the archival plates, and is taking steps to digitize
entire collections.  The United States Naval Observatory
(USNO), for example, digitized most of the known large
Schmidt telescope plates, such as those taken for the
Palomar Observatory Sky Survey, and has made the digitized
images available on the Web.  A program is under way at
Harvard to digitize their plates, but it will take a
number of years to complete any such project.  Until
then, visiting the plate archives will be the only
method of performing archival plate research.

There are other archival research methods.  Datamining
is an accepted practice today, now that so many catalogs
and observations appear on the World Wide Web.  There are
many talented amateurs with excellent computer skills
than can be utilized to find information.  A datamining
example is the paper by Otero (2004), showing how
accessed data from the All Sky Automated Survey (ASAS,
Pojmanski 2002) can be used to provide new elements
for eclipsing binaries.

A final type of archival research is the use of the
AAVSO database for studying long-term behavior of
stars as mentioned earlier.  A subset of
this is when a professional has made a single-epoch
observation of a variable, such as a spectra or
infrared magnitude measurement.  They often request
a light curve of a variable covering the observational
window to see where in the light curve cycle of the
star the observation was made.  The archival
observations can also be used to predict future
maxima or minima for scheduling observations.

\section{Campaigns}

Professional observers often find adequate coverage of
a particular event is very hard to accomplish.  For
example, if HST is to obtain a 3-orbit exposure on
a given date of a specific object, that date is often
not known far in advance.  Scheduling may be performed on
a weekly basis, and the observer not told of the exact
date and time until close to the actual observation.
Finding time at a large ground-based observatory with little
lead time is extremely difficult.

There are several ways in which ground-based support
on a campaign can be accomplished.  If the professional
has at their disposal a locally-owned telescope, they
can often do their own scheduling and insert an observation.
Use of a national facility is harder, as a project may
be granted time, but coordination is very difficult
without disrupting the program of another observer.
One can use a queue-scheduled robotic telescope such
as the Faulkes Telescope or the Liverpool Telescope,
but there are few of these available today and an
agreement may not be in place.  Finally, 
collaborators at other facilities may be able to obtain
time on their telescopes.

Whether any of the above methods will work depends on
many factors, not the least of which are when the
space-based observation is to take place, what is the
geographical location of the ground-based observatory,
the instrument mounted on the available telescope,
and the weather conditions.  A long ground-based
campaign can be even worse, in attempting to coordinate
the times and efforts of many observatories to obtain
nearly complete temporal coverage.  Such techniques
have been used with loose collaborations of professional
sites, as with the Whole Earth Telescope (WET; Nather
et al., 1990) and the Whole Earth Blazar Telescope
(WEBT; Villata et al. 2002).

However, not all campaigns necessarily require large
aperture telescopes.  For a recent campaign on
AE Aqr requested by Mauche (2005), for example, ground-based
photometry was desired in support of an extreme multiwavelength
campaign.  AE Aqr is usually around visual magnitude
11.3, so is easy to reach with almost any amateur telescope.
Flaring activity is possible with this highly magnetic
cataclysmic variable, so careful ground-based monitoring
was necessary to connect any activity seen in X-ray
or gamma-ray regimes with those seen at radio wavelengths
with the VLA.  Because of the large number of amateurs
available, 24-hour coverage over several days is easily
possible for such campaigns, even taking into account
geographical location and weather conditions.

Other campaigns require specific instrumentation or
photometric passbands.  Some professional observatories,
for example, use infrared instruments during bright
time and cannot do optical imaging on those nights.

Some professionals have formed their own campaign
teams of amateurs, for specific programs.  An
example is the Center for Backyard Astrophysics (CBA),
run by J. Patterson of Columbia University.  It
is a team of several dozen amateurs and
small-college professors that monitor cataclysmic
variables, obtaining detailed time-series coverage
when important systems go into outburst.

One of the large advantages of working with an
organization such as the AAVSO is that they can act
as facilitators for campaigns.  The professional
presents a proposal to the AAVSO. The AAVSO advertises
the campaign to their membership, usually including
information on how to take observations to make them
most useful to the requesting professional.  As the
data are uploaded to the AAVSO database, they are
passed on to the professional.  This is much like
having a personal robotic telescope, without the
maintenance and processing headaches!

\section{Outside Expertise}

Not all pro-am collaborations deal with observing.
The usual definition of an amateur is one who does
an activity for the love of it and is not paid,
earning his/her living in some other field.  This
means that amateurs come from all walks of life,
from social workers to CEOs of major corporations.
Often there will be an amateur that has a particular
expertise that is of value for a project.  An
example is someone who works full-time as a software
database expert.  If that person has an interest
in astronomy, they obviously can be of great use
in designing a new database to hold the results of
a major campaign.  Another amateur may be an
electronics engineer, a statistician, a professional
writer or an optician.  Others may, in fact, be
retired professional astronomers.  All of these
amateurs can be utilized if they can be identified.
That is one of the values of a large organization
such as the AAVSO - acting as a matchmaker since
occupation information is often part of a membership
application, or is known internally.

\section{Data Quality}

The AAVSO works diligently to improve the photometric
quality of the observations submitted to its database.
The new charts and sequences that are being produced
eliminate that potential source of error, so that
only the observer's estimation error remains.
Workshops to demonstrate better techniques of visual
observing have been held, with all lectures videotaped
and in process of editing and incorporation into
instructional DVDs.
Workshops on CCD observing, along with manuals, have
been given and produced.  Email lists are available
for observers to ask questions, with experts on call
to give support.  In other words, one of the roles
of a formal amateur organization is training, improving
the skills of the observers.

\begin{figure}[!ht]
\centerline{\hbox{\psfig{figure=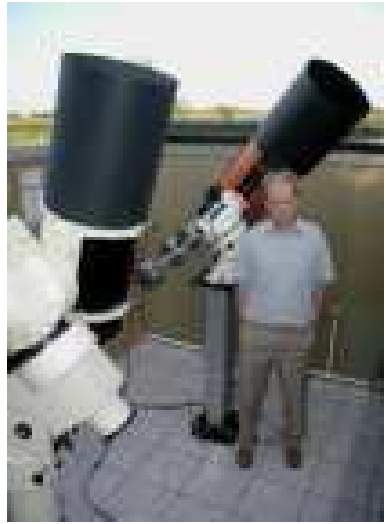,angle=0,clip=,width=5cm}}}
\caption[]{T. Vanmunster and Observatory.}
\label{fig:tonny}  
\end{figure}

\begin{figure}[!ht]
\centerline{\hbox{\psfig{figure=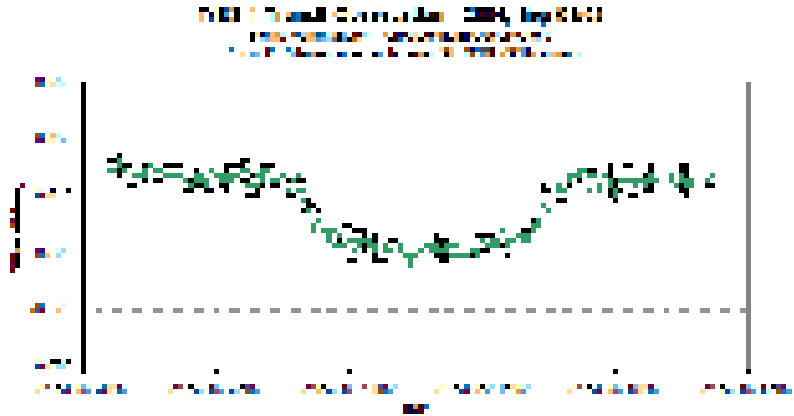,angle=0,clip=,width=11cm}}}
\caption[]{TrES-1 transit as observed by Vanmunster.}
\label{fig:tres1}  
\end{figure}

However, there are many amateurs now that have sufficient
expertise to rival most professionals.  One distinct
advantage amateurs have is that they know their equipment
very well, and have optimized their observing strategies
to account for equipment features and sky conditions.
An example of the quality of existing observations is
from T. Vanmunster, whose twin 35cm telescopes are
shown in Fig.~\ref{fig:tonny}  While Vanmunster concentrates on
cataclysmic variables, he has also been active in
the study of exoplanet transits.  One of his light curves
for the variable TrES-1 is shown in Fig.~\ref{fig:tres1}  This
light curve has less than 5mmag error, showing the
20mmag transit clearly.

\section{Summary}

There are numerous advantages to working with amateurs
in a pro-am collaboration.  The distinction between professional
and amateur has blurred, so that everyone should be treated
as a valued collaborator and given equal footing.
Amateurs are enthusiastic and have the sheer numbers to
provide geographic and temporal coverage for many events.
They tend to be competent, understanding their equipment
better than most professionals visiting national facilities.
They often have other useful skills such as database management.
Many granting agencies require Broader Impact sections in
proposals, showing how research will benefit the rest of
the community.  Including amateurs on your proposals often
helps in satisfying this requirement.

At the same time, working with amateurs is not the same
as working with professional colleagues.  Not all
amateurs are equally experienced, so training may be
required.  While some amateurs may live on isolated
mountaintops, others are in severely polluted urban
environments; the level of equipment and sites will
vary greatly.  Projects may need to be tailored to suit
the abilities of the observers.

Communication is essential.  Amateurs are curious and want
to have the science behind the project explained in simple
terms if possible.  They want to see results and to see that
their observations are not disappearing into a ``black hole."
Try to explain what you want from the amateurs.  Their
only rewards are appreciative comments from the professional,
plus acknowledgements and inclusion as coauthors on papers.

Remember that amateur astronomers are not externally funded.
Whenever possible, include upgrades to their equipment as
part of your proposal.

Training, and finding amateurs for projects, are exactly
what facilitators like the AAVSO are there to perform.
Use their advice and work with them to create campaigns
that utilize the amateur resource properly.  Pro-am
collaborations will further almost anyone's scientific
goals, and can be a rewarding experience.

{}


\begin{thebibliography}{}

\bibitem[Alnso et al. (2004)]{alnso2004}
Alnso, R. et al. 2004, ApJL 613, 153.
\bibitem[Bedding et al. (2002)]{bedding2002}
Bedding, T. R. et al. 2002, MNRAS 337, 79.
\bibitem[Berry et al. (1994)]{berry1994}
Berry, R., Kanto, V., Munger, J. 1994, The CCD Camera Cookbook
(Richmond: Willmann-Bell).
\bibitem[Cornell (1975)]{cornell1975}
Cornell, J. 1975, Sky and Telescope, 50, 160.
\bibitem[Evans (2005)]{evans2005}
Evans, R. 2005, IAUC 8580.
\bibitem[Hearnshaw \& Cottrell (1986)]{hearnshaw1986}
Hearnshaw, J. B., Cottrell, P. L. 1986, Instrumentation and
Research Programmes for Small Telescopes, IAU Symposium 118
(Dordrecht: D. Reidel).
\bibitem[Kaiser (1995)]{kaiser1995}
Kaiser, D. H. 1995, JAAVSO 23, 135.
\bibitem[Kato (2004)]{kato2004}
Kato, T. et al. 2004, PASJ 56, S1.
\bibitem[Liller (1999)]{liller1999}
Liller, W. 1999, JAAVSO 27, 44.
\bibitem[Mauche (2005)]{mauche2005}
Mauche, C. 2005, AAVSO Alert Notice 326.
\bibitem[Millis (1996)]{millis1996}
Millis, R. L. 1996, Proceedings of The Role of Small Telescopes in Modern
Astronomy: The First Annual Lowell Observatory Fall Workshop
(Flagstaff: Lowell Observatory).
\bibitem[Nather et al. (1990)]{nather1990}
Nather, R. E., Winget, D. E., Clemens, J. C., Hansen, C. J. and
Hine, B. P., 1990, ApJ 361, 309.
\bibitem[Otero (2004)]{otero2004}
Otero, S. A. 2004, IBVS 5532,1.
\bibitem[Percy (2000)]{percy2000}
Percy, J. R., Wilson, J. B. 2000, Amateur-Professional Partnerships in
Astronomy (San Francisco: ASP Conf. Proc. 220).
\bibitem[Pojmanski (2002)]{pojmanski2002}
Pojmanski, G. 2002, Acta Astronomica 52, 397.
\bibitem[Price (2005]{price2005}
Price, C. A. 2005, Sky and Telescope, accepted.
\bibitem[Smith (1995)]{smith1995}
Smith, R. C. 1995, JBAA 105, 167.
\bibitem[Szkody et al. (2005)]{szkody2005}
Szkody, P. et al. 2005, in AAVSO Alert Notice 318.
\bibitem[Templeton et al. (2005)]{templeton2005}
Templeton, M. R., Mattei, J. A., Willson, L. A. 2005, AJ 130, 776.
\bibitem[Tonkin (2002)]{tonkin2002}
Tonkin, S. F. ed. 2002, Practical Amateur Spectroscopy
(Berlin: Springer).
\bibitem[Turner (2003)]{turner2003}
Turner, D. G. 2003, JAAVSO 31, 160.
\bibitem[Villata et al. (2002)]{villata2002}
Villata, M. et al. 2002, MmSAI 73, 1191.
\bibitem[Wheatley et al. (2003)]{wheatley2003}
Wheatley, P. J., Mauche, C. W., Mattei, J. A. 2003, MNRAS 345,49.
\bibitem[Wood (1963)]{wood1963}
Wood, F. B. 1963, Photoelectric Astronomy for Amateurs,
(New York: Macmillan).
\bibitem[Wood \& Zarro (1981)]{wood1981}
Wood, P. R., Zarro, D. M. 1981, ApJ 247, 247.

\end{thebibliography}
\end{document}